\documentclass[journal]{IEEEtran}

\ifCLASSINFOpdf

\else

\fi
\usepackage[cmex10]{amsmath}
\usepackage{amssymb}
\usepackage{cite}
\usepackage{graphicx}
\usepackage{array,color}
\usepackage{amsmath}
\usepackage{stfloats}
\usepackage{graphicx}
\usepackage{subfigure}
\usepackage{tabularx}
\usepackage{epsfig,epsf,color,balance,cite}
\usepackage{algorithmic}
\usepackage{algorithm}
\usepackage{epstopdf}
\usepackage{bm}
\usepackage{textcomp}

\begin{document}

\title{Energy Efficient Transmission for Multicast Services in MISO Distributed Antenna Systems}

\author{Hong Ren, \IEEEmembership{Student Member, IEEE}, Nan Liu, \IEEEmembership{Member, IEEE}, Cunhua Pan, \IEEEmembership{Student Member, IEEE}
\thanks{This work is partially supported by the National Basic Research Program of China
(973 Program 2012CB316004), the National Natural Science Foundation of China
under Grants $61571123$ and $61221002$, the Research Fund of National Mobile Communications Research Laboratory, Southeast University (No. 2016A03) and Qing Lan Project.}
\thanks{H. Ren and  N. Liu are with National Mobile Communications Research Laboratory, Southeast University, Nanjing 210096, China. (Email:\{renhong,  nanliu\}@seu.edu.cn).}
\thanks{C. Pan is with the School of Engineering and Digital Arts, University of Kent, Canterbury, Kent, CT2 7NZ, U.K. (Email:{C.Pan@kent.ac.uk}).}
\thanks{Corresponding author: N. Liu.}
}

\maketitle

\maketitle

\vspace{-1cm}\begin{abstract}
This paper aims to solve the energy efficiency (EE) maximization problem for multicast services in a multiple-input single-output (MISO) distributed antenna system (DAS). A novel iterative algorithm is proposed, which consists of solving two subproblems iteratively: the power allocation problem and the beam direction updating problem. The former subproblem can be equivalently transformed into a one-dimension quasi-concave problem that is solved by the golden search method. The latter problem  can be efficiently solved by the existing method. Simulation results show that the proposed algorithm achieves significant EE performance gains over the existing rate maximization method. In addition, when the backhaul power consumption is low, the EE performance of the DAS is better than that of the centralized antenna system (CAS).
\end{abstract}

\vspace{-0.1cm}

\IEEEpeerreviewmaketitle

\vspace{-0.4cm}\section{Introduction}\vspace{-0.15cm}

Recently, distributed antenna system (DAS), large-scale MIMO (LS-MIMO), cloud radio access networks (CRANs) and Heterogeneous networks (Hets) have been regarded as promising techniques to meet the fifth generation (5G) requirements of cellular networks \cite{Andrews2014}. In this paper, our focus is on the DAS. While in the LS-MIMO system, all antennas are deployed in the same geographical location, which causes correlation among the antennas' signals \cite{Wagner2012}, in a DAS, remote access units (RAUs) are located at different places, and each RAU is connected to the central processing unit (CPU) through optical fibers. Through distributed implementation, both the average access distance for the users and the correlations of the antennas can be significantly reduced.  In addition, DAS usually operates on the per-macrocell basis, which means that the RAUs in each macrocell serve the users in its own cell and signals from other macrocells are regarded as interference. Compared with Hets, where interference management is very challenging especially for dense Hets \cite{Insoo2013}, in the DAS, the interference from different RAUs in the same macrocell can be efficiently handled due to the fact that different RAUs in each macrocell cooperate with each other for transmission. Compared with the small-scale interference coordination offered by the DAS, large-scale interference coordination is possible in CRANs, where all base stations (BSs) from a large geographical area are connected to the same computing center via fronthaul links, and base-band processing for the multiple BSs are done jointly at the computing center to cancel inter-cell interference. Some technical challenges for CRAN include the large delays on the fronthaul links and the large amount of overhead and information to enable large-scale interference cancelation \cite{Mugen2014}.

On the other hand, energy efficiency (EE), measured in bit/Hz/Joule, has attracted extensive attention \cite{Hasan2011}, and will become  one of the main concerns in 5G networks \cite{Andrews2014}. Most of the existing papers focus on the EE design for unicast services in DAS \cite{Jingon2013,Jingon2014,Chunlong2014}.
With the increasing demand for video conferences and online games, multicast services should be supported by future networks. Unlike the unicast services,  in a multicast system, all users receive the same service data, and therefore,  the data rate is determined by the user with the worst channel gains. Simulation results in \cite{Boal2007} showed that the DAS can offer a uniform rate distribution over the cell region. Existing EE design methods proposed for unicast services are no longer applicable for multicast services, and there is no study on the EE design for multicast services in DAS.

In this paper, we consider the EE maximization problem for the MISO multicast DAS.  A novel iterative algorithm is proposed to solve the EE maximization problem. In each iteration, two subproblems are solved: the power allocation problem and the beam direction updating problem. By constructing a one-dimension optimization problem that is equivalent to the first subproblem, we utilize the golden search method to find the solution. The second subproblem is quadratically quadratic programming (QCQP). We adopt an existing technique \cite{luozhiquan-08-tsp} to solve this subproblem efficiently. We also analyze the convergence and complexity of the iterative algorithm. Simulation results show that the proposed algorithm outperforms the existing methods in terms of the EE performance.

\vspace{-0.4cm}\section{System Model}\label{systemmodel}\vspace{-0.1cm}

\begin{figure}
\centering
\includegraphics[width=1.815in]{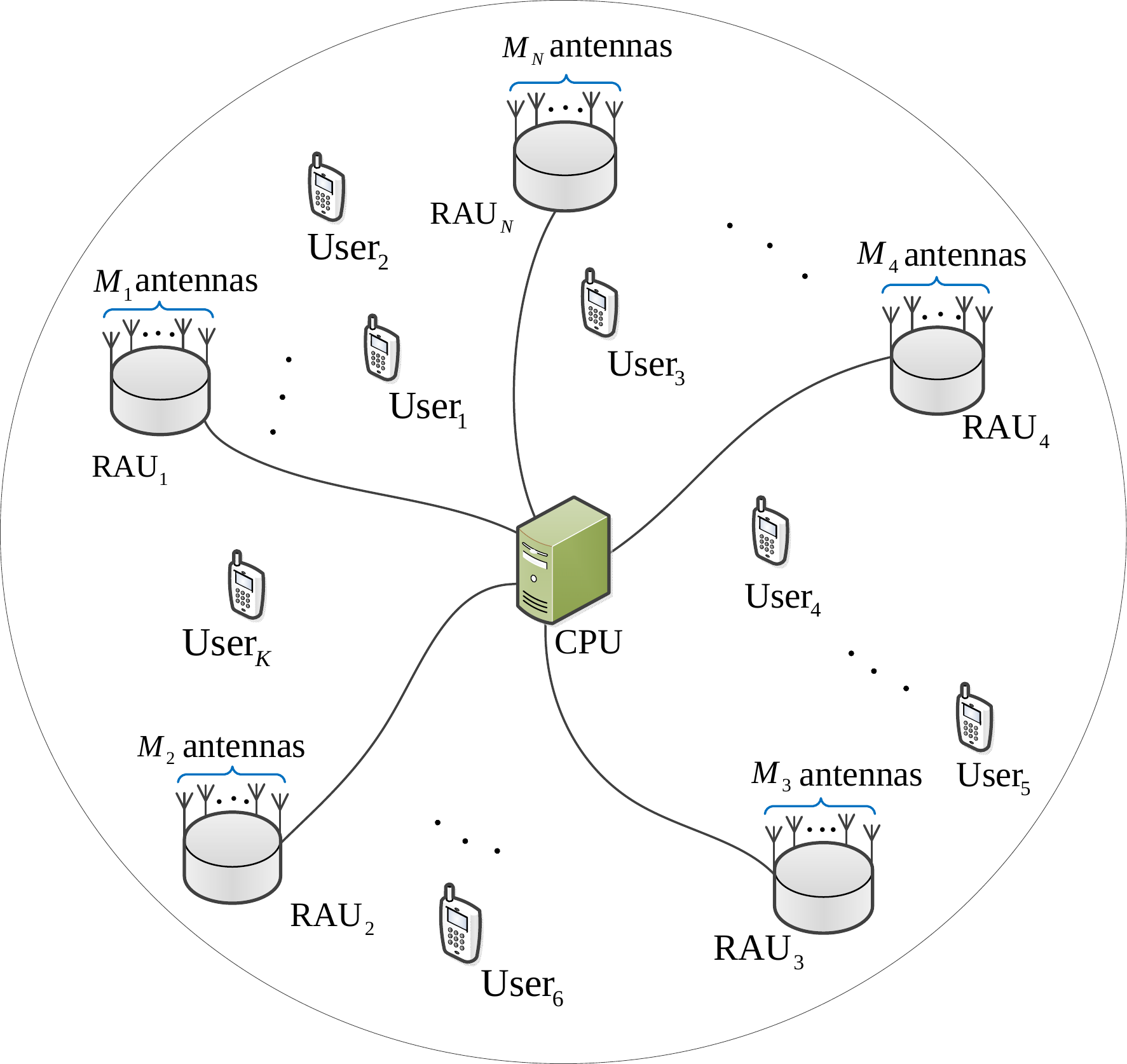}
\caption{Structure of one multicast DAS with $N$ RAUs and $K$ users.}
\label{system}\vspace{-0.7cm}
\end{figure}

We consider a  multicast DAS with $N$ RAUs and $K$ single-antenna users, where RAU $n$ is equipped with ${M_n}$  antennas, $n=1,2,\ldots,N$, as shown in Fig.\ref{system}. All users receive the same data from all the RAUs. Assume that all RAUs are connected to the central processing unit (CPU) through the high speed fiber cable and all RAUs are fully controlled by the CPU. Let $\mathcal{N}=\{1,\cdots,N\}$, $\mathcal{K}=\{1,\cdots,K\}$ be the sets of RAUs and users, respectively. The received signal of the $k$th user is
\vspace{-0.2cm}
\begin{equation}
y_k = \sum\limits_{n \in {\mathcal{N}}} {\bf{h}}_{n,k}^H {\bf{w}}_n x  + {z_k},\forall k \in {\mathcal{K}},\vspace{-0.2cm}
\end{equation}
where ${\bf{h}}_{n,k}\in {\mathbb{C}}^{M_n\times 1}$ denotes the channel vector from  RAU $n$ to the $k$th user, $x\in \mathbb{C}$ is the information symbol for all users in the DAS system with $\mathbb{E}\{x\}=0$ and $\mathbb{E}\{|x|^2\}=1$, ${\bf{w}}_n \in {\mathbb{C}}^{M_n\times 1}$ is the multicast beam-vector at  RAU $n$, $z_k$ is a zero-mean circularly symmetric complex Gaussian noise with variance ${\sigma}^2_k$. The beam-vector ${\bf{w}}_n$ is further factorized to ${\bf{w}}_n = \sqrt{p_n} {\bf{\hat w}}_n$, where ${\bf{\hat w}}_n$ is the beam direction which is normalized to unity, and $p_n$ is the corresponding power. The signal-to-interference-plus-noise ratio (SINR) of user $k$ is
\vspace{-0.1cm}
\begin{equation}\label{SINRk}
{\rm{SIN}}{{\rm{R}}_k} = \frac{{\sum\limits_{n \in {\mathcal{N}}} {{p_n}{{\left| {{\bf{\hat w}}_n^H{{\bf{h}}_{n,k}}} \right|}^2}} }}{{\sigma _k^2}}.\vspace{-0.1cm}
\end{equation}
Then, the worst-case user rate (bit/s/Hz) $R$ is \cite{Tian2012}
\vspace{-0.17cm}
\begin{equation}
R = {\log _2}\left( {1 + \mathop {\min }\limits_{k \in \mathcal{K}} {\rm{SIN}}{{\rm{R}}_k}} \right).
\vspace{-0.17cm}
\end{equation}

To consider the EE design, the total power consumption should be considered, which can be modeled as \cite{Onireti}
\vspace{-0.1cm}
\begin{equation}
\begin{array}{l}
 {P_{\rm{total}}} = \sum\limits_{n \in {\mathcal{N}}} {p_n}+{P_C}+NP_{\rm{bh}},\vspace{-0.1cm}
\end{array}
\end{equation}
where $P_{\rm{bh}}$ represents the backhaul power consumption on each backhaul link\cite{Onireti}, which is used to transmit data and control information, and ${P_C}$ denotes the total circuit power consumption, and is defined as $P_C= \sum\nolimits_{n \in {\mathcal{N}}} {{M_n}{p_c}}  + N{p_0}$  where $p_c$ denotes the circuit power consumption per antenna and $p_0$ denotes the static power consumed at each RAU.

The EE of the multicast DAS system is defined as the ratio of the worst-case user rate to the total power consumption, measured in bit/Hz/Joule. Our objective is to optimize all the beam-vectors at the RAUs to maximize the EE of the DAS system, subject to both the per-RAU power constraints and the worst-case user rate requirement.  The EE maximization problem can be formulated as
\vspace{-0.1cm}
\begin{equation*}\label{originalproblem}
\begin{array}{l}
({\rm{P1}})\quad\  \mathop {\max }\limits_{\{{\bf{\hat w}}_n, p_n, \forall n\}} \frac{R}{P_{\rm{total}}}\\
\qquad\qquad\quad{\rm{s.t.}}\quad {p_n} \le P_n^{\max },\forall n \in {\mathcal{N}},\\
\qquad\qquad\qquad\quad\; R \ge {R_{\min }},\vspace{-0.1cm}
\end{array}
\end{equation*}
where  $R_{\min}$ denotes the minimum rate requirement and $P_n^{\max }$ denotes the transmit power constraint for RAU $n$.

\vspace{-0.3cm}\section{One Novel Algorithm to Solve Problem (P1)}

In this section, we focus on the solution of the original problem (P1). Since (P1) is not a convex problem, it's hard to find its optimal solution. To solve it, we propose a novel iterative algorithm with two steps.

In the first step, the power allocation solution on the beam-vectors is solved with fixed beam directions. By introducing an auxiliary variable $t$, Problem (P1) can be reformulated as
\vspace{-0.1cm}
\begin{equation*}
\begin{array}{l}
({\rm{P2}}) \mathop {\max }\limits_{\{t,p_n, \forall n\}} \frac{{{{\log }_2}\left( {1 + t} \right)}}{{\sum\limits_{n \in {\mathcal{N}}} {{p_n}}  + {P_c}}}\\
\qquad\quad\; {\rm{s.t.}}\quad {\rm{C1:}}\ {p_n} \le P_n^{\max },\;\forall n \in {\mathcal{N}},\\
\qquad\qquad\quad\;\; {\rm{C2:}}\frac{\sum\limits_{n \in {\mathcal{N}}} {{p_n}{g_{n,k}}} }{\sigma _k^2} \ge {\rm{max}}\{2^{R_{\min}} - 1, t\},\forall k \in {\mathcal{K}},
\end{array}
\end{equation*}
where $g_{n,k}={\left| {{\bf{\hat w}}_n^H{{\bf{h}}_{n,k}}} \right|}^2, \forall n,k$. This problem will be solved in Subsection \ref{optimalpower}.  Note that the optimal $t$ is no less than $2^{R_{\min}}-1$.

In the second step, the beam directions are optimized to minimize the total transmit power with fixed SINR $t$ that is  obtained from the first step. The optimization problem is formulated as
\vspace{-0.1cm}\begin{equation*}\label{powermin}
\begin{array}{l}
({\rm{P3}})\quad\  \mathop {\min }\limits_{\{{\bf{w}}_n, \forall n\}}  \sum\limits_{n \in {\mathcal{N}}} {\left\| {{\bf{w}}_n} \right\|_2^2}  \\
\qquad\qquad\;\; {\rm{s}}{\rm{.t}}{\rm{. }}\ \frac{\sum\limits_{n \in {\mathcal{N}}} {{{\left| {{\bf{w}}_n^H{{\bf{h}}_{n,k}}} \right|}^2}} }{{\sigma _k^2}} \ge t,\forall k \in {\mathcal{K}},\\
\qquad\qquad\qquad\;\; \left\| {{\bf{w}}_n} \right\|_2^2 \le P_n^{\max }, \forall n \in {\mathcal{N}}.
\end{array}
\end{equation*}
This problem will be discussed in Subsection \ref{beamdire}. Since the $t$ obtained from the first step is no less than $2^{R_{\min}}-1$, the solution of Problem (P3) satisfies the minimum rate requirement.

By iteratively solving Problem (P2) and Problem (P3), the algorithm that solves Problem (P1), named as the EETM algorithm, is given in Algorithm \ref{algorithmP1}, where ${{\rm{EE}}\left\{ {p_n^{(l )},{\bf{\hat w}}_n^{(l)},\forall n} \right\}}$ denotes the value of the objective function of Problem (P1) when $ p_n=p_n^{(l )},{\bf{\hat w}}_n={\bf{\hat w}}_n^{(l)},\forall n $.

\vspace{-0.2cm}\begin{algorithm}
\caption{Energy Efficient Multicast Transmission (EEMT)}\label{algorithmP1}
\begin{enumerate}
  \item  Initialize $l=1$, accuracy $\delta$, beam directions $\{{\bf{\hat w}}_n^{(0)}, \forall n\}$ and  power allocation $\{p_n^{(0)}, \forall n\}$.
  \item Given beam directions $\{{\bf{\hat w}}_n^{(l-1)}, \forall n\}$, update the power allocation by solving problem (P2) detailed in Algorithm  \ref{powerallalg}. Denote the solution by $\{t^*,p_n^*, \forall n\}$. Let $t^{(l)}=t^{*}$.
  \item Solve problem (P3) via the method in \cite{luozhiquan-08-tsp} where $t=t^{(l)}$. Denote the solution by ${\bf{w}}_n^*,\forall n$. Update  $p_n^{(l)}={\left\| {{\bf{w}}_n^*} \right\|^2}$ and ${\bf{\hat w}}_n^{(l)} = {{{\bf{w}}_n^*} \mathord{\left/
 {\vphantom {{{\bf{w}}_n^*} {\left\| {{\bf{w}}_n^*} \right\|}}} \right.
 \kern-\nulldelimiterspace} {\left\| {{\bf{w}}_n^*} \right\|}}, \forall n$.
  \item If $\left| {{\rm{EE}}\left\{ {p_n^{(l)},{\bf{\hat w}}_n^{(l)},\forall n} \right\} - {\rm{EE}}\left\{ {p_n^{(l - 1)},{\bf{\hat w}}_n^{(l - 1)},\forall n} \right\}} \right| \le \delta $,  output $\{p_n^{(l)},{\bf{\hat w}}_n^{(l)},\forall n\}$. Otherwise, let $l=l+1$ and go to step 2).
\end{enumerate}
\end{algorithm}

\vspace{-0.5cm}\subsection{Algorithm to Solve Problem (P2)}\label{optimalpower}\vspace{-0.1cm}

In this subsection, we present an algorithm to solve Problem (P2) that will be used in step 2) of the EEMT algorithm. Obviously, Problem (P2) is non-convex, so it is difficult to solve it directly. In the following, we construct a tractable optimization problem that is equivalent to Problem (P2).

Define function $f(t)$ as
\vspace{-0.2cm}\begin{equation}\label{powerfun}
\begin{array}{l}
f\left( t \right) = \mathop {\min }\limits_{\{p_n,\forall n\}} \;\sum\nolimits_{n = 1}^N {{p_n}} \\
\qquad\qquad {\rm{s.t.}}\quad  {\rm{C1}},{\rm{C2}}.
\end{array}\vspace{-0.2cm}
\end{equation}
Note that though the problem in (\ref{powerfun}) and Problem (P3) look similar, they are not equivalent. The optimizing variables for the problem in (\ref{powerfun}) are real scalars ${\{p_n, n \in \mathcal{N}\}}$, while that for Problem (P3) are complex vectors $\left\{ {{{\bf{w}}_n},\forall n} \right\}$. The problem in (\ref{powerfun}) is a linear programming problem and the globally optimal solution can be efficiently obtained. The following lemma shows the convexity of $f(t)$.

\textbf{Lemma 1}:
$f(t)$ is a convex function of $t$.

\emph{Proof}: Denote power allocation $\{p_n, \forall n\}$ as an $N$-vector  ${\bf{P}}$, i.e., ${\bf{P}} = {[{p_1}, \cdots ,{p_N}]^T}$. Let ${\bf{P}}_1^\star$, and ${\bf{P}}_2^\star$ be the optimal solution to the problem in (\ref{powerfun}) with $t=t_1$ and $t=t_2 $,  respectively.  Let ${\bf{P}}_3^\star$ be the  optimal solution to the problem in (\ref{powerfun}) with $t=t_3\triangleq vt_1+(1-v)t_2$, for any $0\leq v \leq1$. Construct a power allocation vector ${{\bf{P}}_3} = v{\bf{P}}_1^\star + (1 - v){\bf{P}}_2^\star$.  ${{\bf{P}}_3}$ is a feasible solution of the problem in (\ref{powerfun}) for $t=t_3$ due to the linearity of the constraints ${\rm{C1}},{\rm{C2}}$ in terms of $\{p_n, \forall n\}$ for a given $t$. Hence, we have
\vspace{-0.3cm}\[\begin{array}{l}
vf({t_1}) + (1 - v)f({t_2}) = v\sum\limits_{n = 1}^N {p_{n,1}^*}  + (1 - v)\sum\limits_{n = 1}^N {p_{n,2}^\star} \\
= \sum\limits_{n = 1}^N {{p_{n,3}}}  \ge \sum\limits_{n = 1}^N {p_{n,3}^\star}  = f({t_3})=f(vt_1+(1-v)t_2),
\end{array}\vspace{-0.2cm}\]
where the inequality follows since ${\bf{P}}_3^\star$ is the optimal solution when  $t=t_3$. Therefore, $f(t)$ is a convex function of $t$. \hfill $\Box$

Then, we construct the following problem
\vspace{-0.1cm}
\begin{equation*}
\begin{array}{l}
\left( {\rm{P4}} \right)\ \mathop {\max }\limits_{t \in \left[ {{t_{\min}},{t_{\max }}} \right]} \frac{{{{\log }_2}\left( {1 + t} \right)}}{{f\left( t \right) + {P_C}}},
\end{array}\vspace{-0.1cm}
\end{equation*}
where  ${t_{\min }} = {2^{R_{\min}}} - 1$, ${t_{\max }} = \mathop {\min }\limits_{k \in \mathcal{K}} \ \sum\limits_{n \in {\mathcal{N}}} {{{{g_{n,k}}P_n^{\max }} \mathord{\left/
 {\vphantom {{{g_{n,k}}P_n^{\max }} {\sigma _k^2}}} \right.
 \kern-\nulldelimiterspace} {\sigma _k^2}}} $.

According to Lemma 1, $f(t)$ is a convex function of $t$. Moreover, $\text{log}_2(1+t)$ is a strictly concave function of $t$. Hence, the objective function of (P4) is a strictly quasi-concave function of $t$ \cite{Boyd2004}. Then Problem (P4) has a unique globally optimal solution \cite{Boyd2004}. The following lemma establishes the equivalence between Problem (P4) and Problem (P2).

 \textbf{Lemma 2}: Denote the globally optimal solution to Problem (P4) as $t^{*}$. Let $\{p_n^*, \forall n\}$ be the power profile for achieving $f(t^*)$. Then, the solution $\{t^*, p_n^*, \forall n\}$  is a globally optimal solution to Problem (P2).

\emph{Proof}:  We prove this by contradiction. Suppose that $\{t^\prime, p_n^\prime, \forall n\}$  is a globally optimal solution to Problem (P2), and has a higher EE value than $\{t^*, p_n^*, \forall n\}$. Given $t^\prime$, the power allocation $\{p_n^\prime, \forall n\}$ satisfies the constraints in problem (\ref{powerfun}). Hence, $f(t^\prime)\leq \sum\nolimits_{n \in {\mathcal{N}}} {{{p'}_n}} $ holds. Then, we have the following chain of inequalities
\vspace{-0.1cm}
\begin{equation}
\begin{array}{l}\label{conflict}
\frac{{{\log }_2}\left( {1 + t'} \right)}{{f\left( {t'} \right) + {P_C}}}\! \overset{(a)} \ge \! \frac{{{{\log }_2}\left( {1 + t'} \right)}}{\sum\limits_{n = 1}^N {{p'}_n}  + {P_C}}
 \!\overset{(b)} >\! \frac{{{{\log }_2}\left( {1 + {t^*}} \right)}}{\sum\limits_{n = 1}^N {p_n^*}  + {P_C}}
 \!=\! \frac{{{{\log }_2}\left( {1 + {t^*}} \right)}}{{f\left( {{t^*}} \right) + {P_C}}}\vspace{-0.1cm}
\end{array}
\end{equation}
where $(a)$ holds due to the fact that $f(t^\prime)\leq \sum\nolimits_{n \in {\mathcal{N}}} {{{p'}_n}} $, $(b)$ is due to the hypothesis. Hence,  $t^\prime$ is better than $t^*$, which contradicts the fact that $t^*$ is the globally optimal solution of Problem (P4). Thus, the lemma follows.  \hfill $\Box$

 Since Problem (P4) is strictly quasi-concave,  the golden section search method \cite{kahaner1989numerical} can be used to find the optimal $t^*$. Then, $\{p_n^*, \forall n\}$ can be obtained by solving the linear programming problem (\ref{powerfun}) with $t=t^*$. The algorithm to find the optimal $t^*$ of Problem (P4) and the corresponding $\{p_n^*, \forall n\}$ is given in Algorithm \ref{powerallalg}, the solution of which is the optimal solution to Problem (P2).
\begin{algorithm}
\caption{The golden search algorithm}\label{powerallalg}
\begin{enumerate}
  \item Initialize $a=t_ {\text{min}}$, $b=t_{\text{max}}$, the desired accuracy $\varepsilon$.
  \item Update $t_1=a+0.382(b-a)$ and $t_2=a+0.618(b-a)$.
  \item Obtain $f\left( t_1 \right)$ and $f\left( t_2 \right)$ by solving problem (\ref{powerfun}). If $\frac{{{{\log }_2}\left( {1 + t_1} \right)}}{{f\left( t_1 \right) + {P_C}}}>\frac{{{{\log }_2}\left( {1 + t_2} \right)}}{{f\left( t_2 \right) + {P_C}}}$, $b=t_2$; else $a=t_1$.
  \item If $b-a<\varepsilon$, terminate. Denote the optimal $t$ as $t^*=(t_1+t_2)/2$, and solve problem (\ref{powerfun}) with $t=t^*$. Denote the corresponding optimal $p_n$ as $p_n^*(t^*)$.
       Output $t^*$ and $p_n^*(t^*),\forall n$. Otherwise, go to step 2).
\end{enumerate}
\end{algorithm}
\vspace{-0.2cm}

\vspace{-0.5cm} \subsection{Algorithm to Solve Problem (P3)}\label{beamdire}\vspace{-0.1cm}

Obviously, Problem (P3) is a QCQP problem and the first set of the constraints are non-convex. As proved in \cite{luozhiquan-08-tsp}, Problem (P3) is NP-hard. Similar to  \cite{luozhiquan-08-tsp}, we use the SDP relaxation and Gaussian randomization method to solve Problem (P3).

\vspace{-0.4cm} \subsection{Convergence Analysis of the EETM Algorithm}\vspace{-0.1cm}

\textbf{Lemma 3}: The sequence generated by the EETM Algorithm  always converges.

\emph{Proof}: We show that the objective value of Problem (P1)  monotonically increases during the iterative process, i.e.,  ${\rm{EE}}\left\{ {p_n^{(l)},{\bf{\hat w}}_n^{(l)},\forall n} \right\} \geq {\rm{EE}}\left\{ {p_n^{(l - 1)},{\bf{\hat w}}_n^{(l - 1)},\forall n} \right\}$. Specifically, after solving Problem (P2) in step 2) of the $l$-th iteration, we obtain the new solution  $\{t^*,p_n^*, \forall n\}$. The objective value at this step, i.e., ${\rm{EE}}\left\{ {p_n^*,{\bf{\hat w}}_n^{(l - 1)},\forall n} \right\}$, will be no less than ${\rm{EE}}\left\{ {p_n^{(l - 1)},{\bf{\hat w}}_n^{(l - 1)},\forall n} \right\}$ since $\{p_n^{(l - 1)}, t^{(l - 1)},\forall n\}$ is just the feasible solution of Problem (P2) in step 2) of the $l$-th iteration. After solving Problem (P3) in step 3) of the $l$-th iteration, we obtain the solution of beam-vectors, i.e., ${\bf{w}}_n^*,\forall n$. Let $p_n^{(l)}={\left\| {{\bf{w}}_n^*} \right\|^2}$ and ${\bf{\hat w}}_n^{(l)} = {{{\bf{w}}_n^*} \mathord{\left/
 {\vphantom {{{\bf{w}}_n^*} {\left\| {{\bf{w}}_n^*} \right\|}}} \right.
 \kern-\nulldelimiterspace} {\left\| {{\bf{w}}_n^*} \right\|}}, \forall n$. Note that $\{{{\bf{w}}_n} = \sqrt {p_n^*} {\bf{\hat w}}_n^{(l - 1)}, \forall n\}$ is also feasible for Problem (P3).  According to  \cite{luozhiquan-08-tsp},  $\sum\nolimits_{n = 1}^N {p_n^{(l)}}$ is guaranteed to be no larger than $\sum\nolimits_{n = 1}^N {p_n^*}$. Moreover, the solution $\{p_n^{(l)}, {\bf{\hat w}}_n^{(l)}, \forall n \}$ achieves the achievable rate no less than ${\log _2}(1 + {t^{(l)}})$. Hence,  ${\rm{EE}}\left\{ {p_n^{(l)},{\bf{\hat w}}_n^{(l)},\forall n} \right\}$ is no less than ${\rm{EE}}\left\{ {p_n^*,{\bf{\hat w}}_n^{(l - 1)},\forall n} \right\}$. Hence, we have ${\rm{EE}}\left\{ {p_n^{(l)},{\bf{\hat w}}_n^{(l)},\forall n} \right\} \geq {\rm{EE}}\left\{ {p_n^{(l - 1)},{\bf{\hat w}}_n^{(l - 1)},\forall n} \right\}$.

In addition, the objective value of Problem (P1) has an upper bound, the EETM algorithm converges.  \hfill $\Box$

\vspace{-0.5cm} \subsection{Complexity Analysis of the EETM Algorithm}\vspace{-0.1cm}

In each iteration, two subproblems, i.e., Problems (P2) and (P3), are solved. In the following, we analyze the complexity to solve these two subproblems, respectively.

In each iteration of Algorithm \ref{powerallalg}, the rang of $(b-a)$ will be scaled by 0.618, the above algorithm will stop after ${\left\lceil {{{\log (\varepsilon /({t_{\rm{max}}} - {t_{\rm{min}}}))} \mathord{\left/
 {\vphantom {{\log (\varepsilon /({t_{{\rm{max}}}} - {t_{{\rm{min}}}}))} {\log 0.618}}} \right.
 \kern-\nulldelimiterspace} {\log 0.618}}} \right\rceil }$ iterations, where $\left\lceil \cdot \right\rceil $ denotes the ceiling operator. In each iteration, the computation complexity of the linear programming problem is $\mathcal{O}(N^3(N+2K))$\cite{linear-1989}. Hence, the total complexity to solve Problem (P2) is $\mathcal{O}({\left\lceil {{{\log (\varepsilon /({t_{{\rm{max}}}} - {t_{{\rm{min}}}}))} \mathord{\left/
 {\vphantom {{\log (\varepsilon /({t_{{\rm{max}}}} - {t_{{\rm{min}}}}))} {\log 0.618}}} \right.
 \kern-\nulldelimiterspace} {\log 0.618}}} \right\rceil } (N^3(N+2K)))$.

The complexity of solving Problem (P3) mainly comes from solving the semidefinite programming problem at a complexity cost that is at most $\mathcal{O}((K+N^2)^{3.5})$ \cite{luozhiquan-08-tsp}.

\vspace{-0.435cm}\section{Simulation Results}\vspace{-0.1cm}

In this section, we evaluate the performance of the proposed EETM algorithm. We assume that there are six users  uniformly distributed in a circular cell centered at $(0,0)$ with cell radius set to be $R=1000\ m$. Also, the number of RAUs is assumed to be four, i.e., $N=4$. The location of the $n$-th RAU is  at $\left( {r\cos ({{2\pi (n - 1)} \mathord{\left/
 {\vphantom {{2\pi (n - 1)} N}} \right.
 \kern-\nulldelimiterspace} N}), r\sin ({{2\pi (n - 1)} \mathord{\left/
 {\vphantom {{2\pi (n - 1)} N}} \right.
 \kern-\nulldelimiterspace} N})} \right)$ for $n=1,\cdots, N$, where $r = {{2R\sin ({\pi  \mathord{\left/
 {\vphantom {\pi  N}} \right.
 \kern-\nulldelimiterspace} N})} \mathord{\left/
 {\vphantom {{2R\sin ({\pi  \mathord{\left/
 {\vphantom {\pi  N}} \right.
 \kern-\nulldelimiterspace} N})} {\left( {{{3\pi } \mathord{\left/
 {\vphantom {{3\pi } N}} \right.
 \kern-\nulldelimiterspace} N}} \right)}}} \right.
 \kern-\nulldelimiterspace} {\left( {{{3\pi } \mathord{\left/
 {\vphantom {{3\pi } N}} \right.
 \kern-\nulldelimiterspace} N}} \right)}}$ \cite{Xinzheng2009}.  The distance from each user to each RAU is assumed to be at least $20\ m$. The channel  ${{\bf{h}}_{n,k}}$ is modeled as ${{\bf{h}}_{n,k}} = \sqrt {{\alpha _{n,k}}{S_{n,k}}} {{\bf{\tilde h}}_{n,k}}$, where $\alpha _{n,k}$ denotes the path-loss that is modeled as $38.46 + 35{\log _{10}}(d)$ \cite{Assumptions-2009}, ${S_{n,k}}$ is the log-normal shadow fading  with zero mean and standard deviation 8 dB, ${\bf{\tilde h}}_{n,k}$ represents the small scale fading that is assumed to be zero-mean circularly symmetric complex Gaussian distributed vector with identity covariance matrix. The noise power  is assumed to be  -101 dBm. We assume  $M \triangleq M_n=4$,  $P_n^{\rm{max}}=P_{\rm{max}}, \forall n$, $p_c=29\ {\rm{dBm}}$, $p_0=30\ {\rm{dBm}}$.

\begin{figure}
\begin{minipage}[t]{0.49\linewidth}
\centering
\includegraphics[width=1.75in]{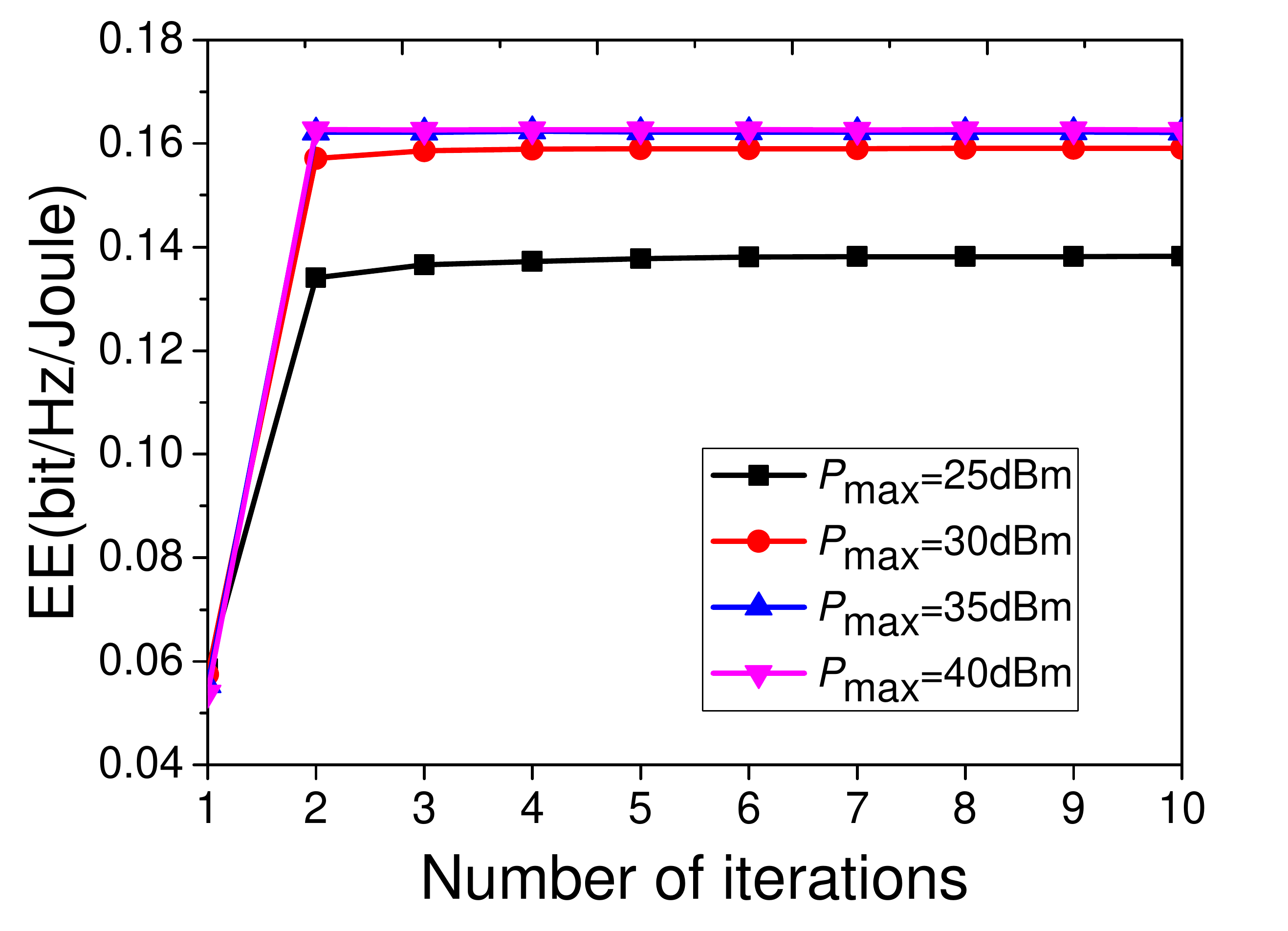}
\caption{Convergence behavior under different power constraints.}\vspace{-0.4cm}
\label{convergence}\vspace{-0.35cm}
\end{minipage}%
\hfill
\begin{minipage}[t]{0.49\linewidth}
\centering
\includegraphics[width=1.75in]{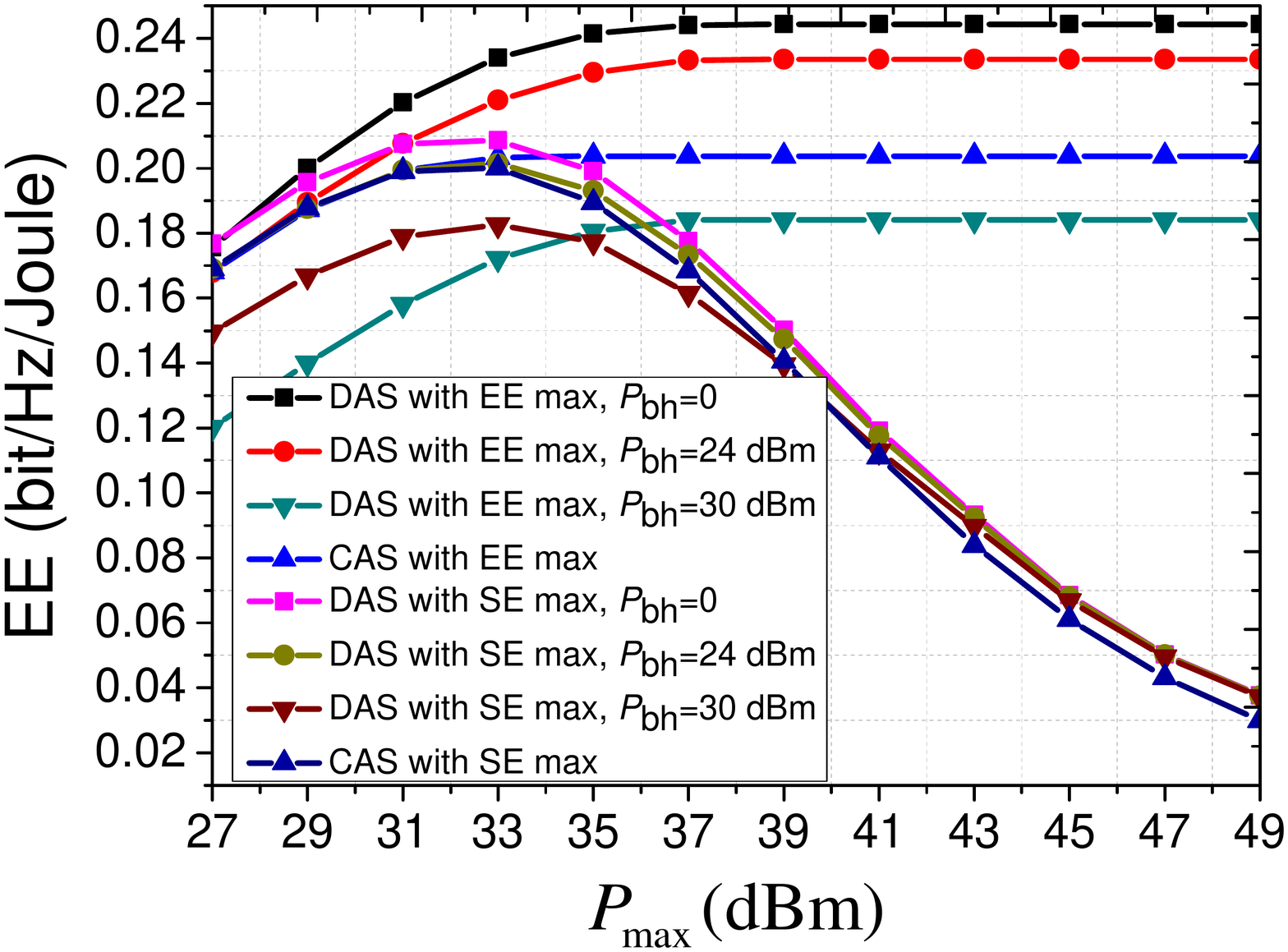}
\caption{EE performance under different power constraints.}\vspace{-0.4cm}
\label{Performancecom}\vspace{-0.35cm}
\end{minipage}

\end{figure}
%
 We first study the convergence behaviour of the EETM algorithm. Fig.~\ref{convergence} illustrates the EE versus the number of iterations of the EETM algorithm under various power constraints when $P_{\rm{bh}}=0$. It is seen from this figure that the EE monotonically increases and converges rapidly, which confirms the theoretical result of Lemma 3. Note that only one iteration is sufficient for the algorithm to converge, which makes our algorithm suitable for practical applications.

We compare our proposed EETM algorithm (label as `DAS with EE Max') with the following methods: the worst-case rate maximization for DAS (labeled as `DAS with Rate Max'), EE maximization for the centralized antenna system (CAS) where all antennas are placed at the center of the cell (label as `CAS with EE Max') and the worst-case rate maximization for CAS (labeled as `CAS with Rate Max'). For fairness, we assume that the maximum transmit power at the CAS is equal to $NP_{\rm{max}}$, the number of antennas at the CAS is $NM$. Moreover, the total circuit power consumption is assumed to be $Np_c+NMp_0$.

Fig.~\ref{Performancecom} shows the EE under different power constraints for various methods. It can be seen from this figure that in the regime of low  power constraint,  the EE achieved by the method aiming for the worst-case rate maximization is almost the same as that achieved by the method aiming for the EE maximization, meaning that full transmit power should be used to achieve the maximum worst-case rate and EE in this regime. However, in the high transmit power regime, the EE achieved by the EE-oriented method stays constant, but the EE corresponding to the worst-case rate-oriented method decreases significantly. This is due to the fact that the increase of the worst-case rate cannot compensate for the negative effect of the increase of the transmit power. As expected, the EE performance of the DAS decreases with the backhaul power consumption parameter $P_{\rm{bh}}$. If $P_{\rm{bh}}$ is low, ignoring signaling overhead and assuming perfect backhaul, the EE performance of the DAS is better than that of the CAS, meaning that to achieve the best EE performance, the antennas should be placed in a distributed way. However, if the parameter $P_{\rm{bh}}$ is large, the EE performance of the DAS is inferior to that of the CAS. A more meaningful comparison between the EE of the DAS and the CAS considering signaling overhead, backhaul availability, capacity and latency etc. is part of the future work.

\vspace{-0.435cm}\section{Conclusion}\label{conclusion}\vspace{-0.1cm}

In this paper, we have studied the EE optimization problem for multicast services in MISO DAS, where both the per-RAU power constraints and the worst-case rate requirement are taken into account. We provide a novel iterative algorithm to solve the original EE optimization problem, which consists of solving two subproblems: the  power allocation problem and the beam direction updating problem. Though the power allocation problem is non-convex, we construct an equivalent problem consisting of a linear programming problem and a quasi-concave problem. The convergence of the iterative algorithm is proved. Simulation results show that our proposed algorithm converges rapidly. Furthermore, when the backhaul power consumption is low, the EE performance of the DAS is better than that of the CAS, indicating that to achieve the best EE performance, the antennas should be placed in a distributed way.



\
\



\vspace{-0.6cm}\bibliographystyle{IEEEtran}
\bibliography{myre}




\end{document}